\input harvmac
\input tables.tex
\noblackbox
\def\ias{\vbox{\sl\centerline{School of Natural Sciences, 
Institute for Advanced Study}%
\centerline{Olden Lane, Princeton, N.J. 08540 USA}}}

\footnotefont

\nref\one{J. Polchinski, Phys. Rev. Lett. 75 (1995) 4724, 
hep-th/9510017.}
\nref\two{A. Sagnotti, in {\it {Cargese '87}}, ``Nonperturbative
Quantum Field Theory'', eds. G.Mack {\it{et al.}} (Pergamon Press,
Oxford, 1988), p.521.}
\nref\three{P. Ho{\u r}ava, Nucl. Phys. {\bf{B327}} (1989) 461\semi
Phys. Lett. {\bf {B231}} (1989) 251.}
\nref\four{J. Dai, R. Leigh, and J. Polchinski, Mod. Phys. Lett.
{\bf{A4}} (1989) 273.}
\nref\five{M. Bianchi and A. Sagnotti, Phys. Lett. {\bf{B247}}
(1990) 517 \semi Nucl. Phys. {\bf{B361}} (1991) 519.}
\nref\six{E. Gimon and J. Polchinski, hep-th/9601038.}
\nref\seven{A. Dabholkar and J. Park, hep-th/9602030.}
\nref\eight {E. Gimon and C. Johnson, hep-th/9604129.}
\nref\nine{A. Dabholkar and J. Park, hep-th/9604178.}
\nref\ten{M. Berkooz and R. Leigh, hep-th/9605049.}
\lref\eighto{D. Morrison and C. Vafa, hep-th/9603161.}
\lref\nineo{A. Sen, hep-th/9605150.}
\lref\teno{E. Gimon and C. Johnson, hep-th/9606176.}
\lref\eleveno{M. Green and J. Schwarz, Phys. Lett. {\bf{B149}}
(1984)117.}
\lref\twelveo{A. Sagnotti, Phys. Lett. {\bf{B294}} (1992)196.}
\lref\pol{J. Polchinski, hep-th/9606165.}
\nref\eleven{J. Schwarz, hep-th/9512053.}
\nref\twelve{E. Witten, hep-th/9507121.}
\nref\thirteen{A. Strominger, hep-th/9512059.}
\nref\fourteen{E. Witten, hep-th/9512219.}
\nref\fifteen{O. Ganor and A. Hanany, hep-th/9602120.}
\nref\sixteen{N. Seiberg and E. Witten, hep-th/9603003.}
\lref\bz{J. Blum and A. Zaffaroni, hep-th/9607019.}
\lref\sen{A. Sen, hep-th/9602010.}
\lref\howi{P. Ho{\u r}ava and E. Witten, Nucl. Phys. 
{\bf{B460}}(1996)506, hep-th/9510209.}
\lref\fth{C. Vafa, hep-th/9602022.}
\lref\vamo{D. Morrison and C. Vafa, hep-th/9602114.}
\lref\many{M. Bershadsky, K. Intriligator, S. Kachru,
D. Morrison, V. Sadov, and C. Vafa; hep-th/9605200.}
\lref\dmw{M. Duff, R. Minasian, and E. Witten; hep-th/9601036.}
\lref\hulto{C. Hull and P. Townsend, Nucl. Phys. {\bf{B438}}(1995)
109, hep-th/9410167.}
\lref\vafwit{C. Vafa and E. Witten, hep-th/9507050.}
\lref\dt{C. Vafa and E. Witten, J. Geom. Phys. 15 (1995) 189,
hep-th/9409188.}
\lref\vois{C. Voisin, in {\it{Journ{\' e}es de G{\' e}ometrie
Alg{\' e}brique d'Orsay}}, {\it{Ast{\' e}risque}} No. 218 (1993) 273.}
\lref\borc{C. Borcea, ``K3 Surfaces with Involution and Mirror
Pairs of Calabi-Yau Manifolds'', to appear in {\it{Essays on
Mirror Manifolds II}}.}
\lref\gomu{R. Gopakumar and S. Mukhi, hep-th/9607057.}
\lref\sentwo{A. Sen, hep-th/9603113.}
\lref\orb{L. Ib{\' a}{\~ n}ez, J. Mas, H. Nilles, and F. Quevedo;
Nucl. Phys. {\bf{B301}} (1988) 157.}
\lref\supone{E. Witten, hep-th/9604030 .}
\lref\suptwo{R. Donagi, A. Grassi, and E. Witten, hep-th/9607091 .}

\Title{\vbox{\baselineskip12pt
\hbox{IASSNS-HEP-96/84}\hbox{hep-th/9608053}}}
{\vbox{\centerline{F Theory Orientifolds, M Theory Orientifolds,}
\centerline{ and Twisted Strings}}}

{\bigskip
\centerline{Julie D. Blum}
\bigskip
\ias

\bigskip
\medskip
\centerline{\bf Abstract}

Orientifolds of the type IIB superstring that descend from F
theory and M theory orbifolds are studied perturbatively.  One
finds strong evidence that a previously ignored twisted open
string is required in these models.  An attempt is made to 
interpret the $J$ type torsion in F theory where one finds a
realization of Gimon and Johnson's models which does not require 
these twisted strings. S-duality also provides evidence for
these strings.

}

\Date{8/96}

\newsec{Introduction}

During the last year the recognition of Dirichlet(D)-branes \one$\,$
as the
sought after Ramond-Ramond solitons in various weak-strong coupling
dualities of string theory has renewed interest in deriving type
I string theories as orientifolds of the type IIB theories \refs
{\two - \ten}.
Six-dimensional string theories, in particular, have received much
attention partly due to a couple of factors.  The massless
spectrum of chiral theories in six dimensions is highly
constrained by anomaly cancellation \eleven , and infrared 
divergences
at singular points in the moduli space of these theories
have been interpreted to signalize the appearance of fascinating
but not well understood tensionless strings \refs{\twelve - \sixteen}.

Studying M and F theories has enlarged the number and furthered
the understanding of string dualities.  The moduli space of
heterotic string compactification to six dimensions has been
described geometrically in F theory \fth\vamo\eighto\many$\,$  and 
more intuitively in
M theory \dmw.  This paper will continue the investigation of \bz$\,$
into orientifolds that correspond to M and F theory models.
In F theory we will study $\bf{Z_2\times Z_n}$ and $\bf{Z_4}$
orbifolds of $\bf{T^6}$.  Notations will be as in \bz .  The
$\bf{Z_2}$ will act on the elliptic fiber and correspond to
$\Omega (-1)^{F_L}R_3$ in the IIB theory.  The $\bf{Z_2\times Z_2}$
case was previously discussed in \bz .  For $n$ even 
we will add discrete
torsion and obtain some other models.  It is important to
emphasize that in Polchinski's notation \pol$\,$ $\Omega$ rather than
$\Omega J$ should be used here.  The details have not been
worked out rigorously, but it appears that models sharing
the same moduli space as that of Gimon and Johnson's
models \eight$\,$ can be achieved in F theory through a $J$ type twist.
There are other 
possible orbifolds involving $\bf{Z_2}$ shifts which will
not be discussed here.\foot{A. Zaffaroni has considered these
models.}  The irreducible anomaly of the massless spectrum will
be exactly zero in the F theory models.  However, we will discover
in IIB string theory that we must introduce a previously 
unnoticed twisted open string to obtain agreement 
of the IIB and F theory
spectrums and to cancel the IIB anomalies.  We will show that the 
IIB tadpole anomaly is proportional to the anomaly of these twisted 
strings.  Interestingly enough, one of the models with discrete
torsion will turn out to be the ubiquitous model of 
\sen\seven\fth .

Next we will find more evidence for these twisted strings by
studying orientifolds of M theory on $\bf{T^5/Z_n}$.  The 
$\bf{Z_2\times Z_n}$ models correspond to the heterotic
string on $K3$ orbifolds or standard type I compactifications (1 tensor
multiplet) on $K3$ \howi\teno\dmw$\,$ and will not be discussed here.  
We will show
that the $\bf{T^5/Z_4}$ and $\bf{T^5/Z_6}$ orbifolds are related by 
T-duality to orientifolds of IIB.  Naively, these orientifolds
appeared to be the $\bf{Z_4}$ and $\bf{Z_6}$ B models of Gimon 
and Johnson \eight$\,$
which did not seem to agree with M theory.  The $SL(2,{\bf{Z}})$
(S)-duality of IIB was expected to interchange $(-1)^{F_L}$ and
$\Omega$ \hulto\vafwit.  However, the massless spectrum of IIB on the
orbifold ${\bf{T^4}}/(-1)^{F_L}{\bf{Z_4}}$ 
did not match that of Gimon and 
Johnson on ${\bf{T^4}}/\Omega {\bf{Z_4}}$.\foot{A. Zaffaroni noticed
this discrepancy.}  We will provide some evidence that the 
resolution of these puzzles is that the above $\Omega$ of Gimon
and Johnson is really $\tilde\Omega=\Omega J$ as discussed in
\teno.  Replacing $\Omega J$ by $\Omega$ will cause twisted
strings to appear in the spectrum and resolve the S-duality
puzzle.  We will also find agreement with the M theory version
of these models resolving the T-duality puzzle.

We will give two pieces of evidence for these twisted strings.
The Klein Bottle contribution of the $\bf{Z_4}$ orientifold
will equal the partition function of S-dual sectors 
corresponding to the $(-1)^{F_L}\bf{Z_4}$ theory when the
modulus of the torus is purely imaginary and equals that of
the Klein bottle.  However, there is one subtlety here that only
affects the massive spectrum:$(-1)^{F_L}$ should be replaced
by $(-1)^{F_L}\Omega _b$ to get full agreement where $\Omega _b$
exchanges only left and right-moving, space-time bosonic
oscillators.  Given that the $(-1)^{F_L}\bf{Z_4}$ orbifold 
requires twisted
closed strings to be a consistent, modular invariant theory
we suspect that the $\Omega\bf{Z_4}$ theory will require a similar
contribution.  There is again a subtlety in this argument
which leads to the second piece of evidence for these strings.
It is possible to calculate a perturbative open string twisted
by $\Omega J \bf{Z_4}$.  This string is analogous to the left-moving
sector of the $(-1)^{F_L}\bf{Z_4}$ twisted string.  Applying
$\Omega$ to this string gives the analogous right-moving sector.
This string has unusual boundary conditions that make sense
for a $\bf{Z_4}$ orbifold.  However, twisting by $\Omega \bf{Z_4}$
gives a string without oscillator modes in the compactified
directions violating supersymmetry since the Ramond(R) and
Neveu-Schwarz(NS) sectors will not match.  Since the twisted
strings should exist in the $\Omega$ but not $\Omega J$ theory,
one suspects that $\Omega$ transforms into $\Omega J$ under the
``modular'' (duality) transformation of the orientifold theory and that
$J$ projects out the unwanted $\Omega$ twisted strings in the
$\Omega J$ theory.  Similar considerations will apply to the
$\bf{Z_6}$ theory.  We do not know whether these strings have a
D-brane interpretation, perhaps by putting enough of them together
to get a $\bf{Z_2}$ twisted object.  Many of these questions 
will not be resolved in this paper.

This paper is organized as follows.  Section two will discuss the
F theory models first from the F theory perspective and then
the IIB perturbative point of view.  Section three will consider
the M theory models showing first the S and T-duality relations
and then providing further evidence for the twisted strings.
Section four will discuss the results.

\newsec{Orientifolds from F Theory}
\subsec{F theory orbifolds}

Results from F theory orbifolds are summarized in the following
table.  The notations are as follows:$-$ is the $\bf{Z_2}$
generator, $i$ is the $\bf{Z_4}$ generator, $\omega$ the
$\bf{Z_6}$ generator, and $\omega ^2$ the $\bf{Z_3}$ generator.
The action of a group element on $\bf{T^6}$ is given and followed
by the Hodge numbers, $h_{11}$ and $h_{21}$, for that sector.
As in \bz$\,$ the coordinates on $\bf{T^6}$ are $z_3$, 
$z_4$, and $z_5$.
Tensors are abbreviated as T, vectors as V, hypermultiplets
as H, discrete torsion as D.T., $J$ will indicate another
type of discrete torsion, and $s$ will stand for sector.  The action
of group elements on the $z_n$ is listed in the appropriate column.

\centerline{\bf{Table 1}:}
\smallskip
\centerline{F Theory Orientifolds}
\smallskip
\begintable
Model|$(h_{11},h_{21})$|T|V|H|$z_5$|$z_3$|$z_4$|
$(h_{11},h_{21})^s$|$T^s$|
$V^s$|$H^s$\crthick
$\bf{Z_2\times Z_2}$|$(51,3)$|$17$|$SO(8)^8$|$4$|$+$|$+$|$+$|
$(3,3)$|$1$|$0$|$4$\cr
|||||$-$|$-$|$+$|$(16,0)$|$0$|$SO(8)^4$|$0$\cr
|||||$-$|$+$|$-$|$(16,0)$|$0$|$SO(8)^4$|$0$\cr
|||||$+$|$-$|$-$|$(16,0)$|$16$|$0$|$0$\crthick
$\bf{Z_4}$|$(31,7)$|$13$|$SO(8)^4$|$8$|$+$|$+$|$+$|$(5,1)$|$3$|$0$|
$2$\cr
|||||$+$|$-$|$-$|$(10,6)$|$10$|$0$|$6$\cr
|||||$-$|$i$|$i$|$(16,0)$|$0$|$SO(8)^4$|0\cr
|||||$-$|$-i$|$-i$|$(0,0)$|||\crthick
$\bf{Z_2\times Z_3}$|$(35,11)$|$13$|$SO(8)^5$|$SO(8)+8$|$+$|$+$|$+$|
(3,1)|$1$|$0$|$2$\cr
|||||$+$|$\omega^2$|$\omega^4$|$(6,3)$|$6$|$0$|$3$\cr
|||||$+$|$\omega^4$|$\omega^2$|$(6,3)$|$6$|$0$|$3$\cr
|||||$-$|$-$|$+$|$(8,4)$|$0$|$SO(8)^2$|$SO(8)$\cr
|||||$-$|$\omega$|$\omega^2$|$(12,0)$|$0$|$SO(8)^3$|$0$\cr
|||||$-$|$\omega^5$|$\omega^4$|$(0,0)$|||\endtable
(table cont'd)
\vfil\eject
(table 1 cont'd)
\begintable
$\bf{Z_2\times Z_4}$|$(61,1)$|$19$|$SO(8)^{10}$|$2$|$+$|$+$|$+$|
$(3,1)$|
$1$|$0$|$2$\cr
|||||$-$|$-$|$+$|$(12,0)$|$0$|$SO(8)^3$|$0$\cr
|||||$-$|$+$|$-$|$(12,0)$|$0$|$SO(8)^3$|$0$\cr
|||||$+$|$-$|$-$|$(10,0)$|$10$|$0$|$0$\cr
|||||$+$|$i$|$-i$|$(4,0)$|$4$|$0$|$0$\cr
|||||$+$|$-i$|$i$|$(4,0)$|$4$|$0$|$0$\cr
|||||$-$|$i$|$i$|$(16,0)$|$0$|$SO(8)^4$|$0$\cr
|||||$-$|$-i$|$-i$|$(0,0)$|||\crthick
$\bf{Z_2\times Z_6}$|$(51,3)$|$17$|$SO(8)^8$|$4$|$+$|$+$|$+$|$(3,1)$|
$1$|$0$|$2$\cr
|||||$-$|$-$|$+$|$(8,0)$|$0$|$SO(8)^2$|$0$\cr
|||||$-$|$+$|$-$|$(8,0)$|$0$|$SO(8)^2$|$0$\cr
|||||$+$|$-$|$-$|$(6,0)$|$6$|$0$|$0$\cr
|||||$+$|$\omega$|$\omega^5$|$(1,0)$|$1$|$0$|$0$\cr
|||||$+$|$\omega^2$|$\omega^4$|$(4,1)$|$4$|$0$|$1$\cr
|||||$+$|$\omega^4$|$\omega^2$|$(4,1)$|$4$|$0$|$1$\cr
|||||$+$|$\omega^5$|$\omega$|$(1,0)$|$1$|$0$|$0$\cr
|||||$-$|$\omega$|$\omega^2$|$(8,0)$|$0$|$SO(8)^2$|$0$\cr
|||||$-$|$\omega^2$|$\omega$|$(8,0)$|$0$|$SO(8)^2$|$0$\cr
|||||$-$|$\omega^4$|$\omega^5$|$(0,0)$|||\cr
|||||$-$|$\omega^5$|$\omega^4$|$(0,0)$|||\crthick
$\bf{Z_2\times Z_2}$,D.T.|$(3,51)$|$1$|$0$|$SO(8)^8+20$|$+$|$+$|$+$|
$(3,3)$|$1$|$0$|$4$\cr
|||||$-$|$-$|$+$|$(0,16)$|$0$|$0$|$SO(8)^4$\cr
|||||$-$|$+$|$-$|$(0,16)$|$0$|$0$|$SO(8)^4$\cr
|||||$+$|$-$|$-$|$(0,16)$|$0$|$0$|$16$\endtable
(table cont'd)
\vfil\eject
(table 1 cont'd)
\begintable
$\bf{Z_2\times Z_4}$,D.T.|$(21,9)$|$11$|$SO(8)^{2}$|$10$|$+$|$+$|$+$|
$(3,1)$|
$1$|$0$|$2$\cr
|||||$-$|$-$|$+$|$(4,0)$|$0$|$SO(8)$|$0$\cr
|||||$-$|$+$|$-$|$(4,0)$|$0$|$SO(8)$|$0$\cr
|||||$+$|$-$|$-$|$(10,0)$|$10$|$0$|$0$\cr
|||||$+$|$i$|$-i~$|$(0,4)$|$0$|$0$|$4$\cr
|||||$+$|$-i~$|$i~$|$(0,4)$|$0$|$0$|$4$\cr
|||||$-$|$i$|$i$|$(0,0)$|||\cr
|||||$-$|$-i~$|$-i~$|$(0,0)$|||\crthick
$\bf{Z_2\times Z_6}$,D.T.|$(19,19)$|$9$|$SO(8)^2$|$SO(8)^2 +12$|
$+$|$+$|$+$|$(3,1)$|
$1$|$0$|$2$\cr
|||||$-$|$-$|$+$|$(0,4)$|$0$|$0$|$SO(8)$\cr
|||||$-$|$+$|$-$|$(0,4)$|$0$|$0$|$SO(8)$\cr
|||||$+$|$-$|$-$|$(0,6)$|$0$|$0$|$6$\cr
|||||$+$|$\omega$|$\omega^5$|$(0,1)$|$0$|$0$|$1$\cr
|||||$+$|$\omega^2$|$\omega^4$|$(4,1)$|$4$|$0$|$1$\cr
|||||$+$|$\omega^4$|$\omega^2$|$(4,1)$|$4$|$0$|$1$\cr
|||||$+$|$\omega^5$|$\omega$|$(0,1)$|$0$|$0$|$1$\cr
|||||$-$|$\omega$|$\omega^2$|$(4,0)$|$0$|$SO(8)$|$0$\cr
|||||$-$|$\omega^2$|$\omega$|$(4,0)$|$0$|$SO(8)$|$0$\cr
|||||$-$|$\omega^4$|$\omega^5$|$(0,0)$|||\cr
|||||$-$|$\omega^5$|$\omega^4$|$(0,0)$|||\crthick
${\bf{Z_2\times Z_3}}, J$|$(20,14)$|$10$|$SO(8)^2$|
$SO(8)+11$|$+$|$+$|$+$|
(3,1)|$1$|$0$|$2$\cr
|||||$+$|$\omega^2$|$\omega^4$|$(3,6)$|$3$|$0$|$6$\cr
|||||$+$|$\omega^4$|$\omega^2$|$(6,3)$|$6$|$0$|$3$\cr
|||||$-$|$-$|$+$|$(8,4)$|$0$|$SO(8)^2$|$SO(8)$\cr
|||||$-$|$\omega$|$\omega^2$|$(0,0)$|||\cr
|||||$-$|$\omega^5$|$\omega^4$|$(0,0)$|||\endtable
(table cont'd)
\vfil\eject
(table 1 cont'd)
\begintable
${\bf{Z_4}}, J$|$(11,11)$|$9$|$0$|$12$|$+$|$+$|$+$|$(5,1)$|
$3$|$0$|$2$\cr
|||||$+$|$-$|$-$|$(6,10)$|$6$|$0$|$10$\cr
|||||$-$|$i$|$i$|$(0,0)$|||\cr
|||||$-$|$-i$|$-i$|$(0,0)$|||\cr
${\bf{Z_2\times Z_4}}, J$|$(7,31)$|$5$|$0$|$16+SO(8)^4$|$+$|
$+$|$+$|$(3,1)$|
$1$|$0$|$2$\cr
|||||$-$|$-$|$+$|$(0,12)$|$0$|$0$|$SO(8)^3$\cr
|||||$-$|$+$|$-$|$(0,4)$|$0$|$0$|$SO(8)$\cr
|||||$+$|$-$|$-$|$(0,10)$|$0$|$0$|$10$\cr
|||||$+$|$i$|$-i$|$(0,4)$|$0$|$0$|$4$\cr
|||||$+$|$-i$|$i$|$(4,0)$|$4$|$0$|$0$\cr
|||||$-$|$i$|$i$|$(0,0)$|||\cr
|||||$-$|$-i$|$-i$|$(0,0)$|||\crthick
${\bf{Z_2\times Z_6}}, J$|$(9,21)$|$7$|$0$|$14+SO(8)^3$|$+$|$+$|
$+$|$(3,1)$|
$1$|$0$|$2$\cr
|||||$-$|$-$|$+$|$(0,4)$|$0$|$0$|$SO(8)$\cr
|||||$-$|$+$|$-$|$(0,4)$|$0$|$0$|$SO(8)$\cr
|||||$+$|$-$|$-$|$(6,0)$|$6$|$0$|$0$\cr
|||||$+$|$\omega$|$\omega^5$|$(0,1)$|$0$|$0$|$1$\cr
|||||$+$|$\omega^2$|$\omega^4$|$(0,5)$|$0$|$0$|$5$\cr
|||||$+$|$\omega^4$|$\omega^2$|$(0,5)$|$0$|$0$|$5$\cr
|||||$+$|$\omega^5$|$\omega$|$(0,1)$|$0$|$0$|$1$\cr
|||||$-$|$\omega$|$\omega^2$|$(0,0)$|||\cr
|||||$-$|$\omega^2$|$\omega$|$(0,0)$|||\cr
|||||$-$|$\omega^4$|$\omega^5$|$(0,0)$|||\cr
|||||$-$|$\omega^5$|$\omega^4$|$(0,0)$|||
\endtable

In the above table a few points need further clarification.  We
have assumed that the gauge group or global symmetry in
these models is always $SO(8)$.  The justification is that
Morrison and Vafa \eighto$\,$ have stated and Sen \nineo$\,$ 
has shown in detail
that a $\bf{Z_2}$ orbifold singularity on the fiber corresponds to
a $D_4$ singularity yielding an $SO(8)$ group.  Discrete torsion
has been described by \dt :In a $\bf{Z_2\times Z_{2n}}$ model
the discrete torsion lies in $\bf{Z_2}$.  Let $a$ be the generator
of $\bf{Z_2}$ and $b$ the generator of $\bf{Z_{2n}}$.  A generic
group element $a^{m_1}b^{n_1}$ in the sector twisted by 
$a^{m_2}b^{n_2}$ receives an extra factor $(-1)^{m_1 n_2-m_2 n_1}$
when one projects in that sector by summing over the action
of group elements.  

The $J$ type of torsion appears to be more complicated but
seems worth understanding since some interesting models are
realized by it.  $J$ can be described as follows.  For the
$\bf{Z_2\times Z_{2n}}$ cases, $\bf{Z_{2n}}$ twists, $t_2$, 
are assigned a value in $\bf{Z_2}$.  Using the same notation
as for the discrete torsion case, the value is $0$ if $n_2=0$, mod 3
and $1$ otherwise.  The values of $p_1\equiv m_1$ and $t_1\equiv m_2$
already lie in $\bf{Z_2}$.  For the $\bf{Z_2\times Z_3}$ case,
$t_2=n_2$, mod 2, and for the $\bf{Z_4}$ case $t_2=1$ for twisted
sectors and $0$ for the untwisted sector.  One defines $p_2=n_1$, mod 2
for the $\bf{Z_2\times Z_6}$ case, $p_2=0$ for the $\bf{Z_2\times Z_3}$
case, and $p_2=n_1\in\bf{Z_4}$ for the other two cases. Then,
$J$ acts by giving an extra $\bf{Z_2}$ or $\bf{Z_4}$ factor to
projectors in the various twisted sectors and with a $\bf{Z_2}$
action on $z_3$.  This can be summarized in the following table.

\centerline{\bf{Table 2}:}

\centerline{$J$ Torsion}
\medskip
\begintable
Model|$J$|Action on $z_3$\crthick
$\bf{Z_2\times Z_3}$|$e^{\pi it_2 p_1}$|$1$\cr
$\bf{Z_4}$|$e^{\pi i t_2 p_2}$|$1$\cr
$\bf{Z_2\times Z_4}$|$e^{\pi it_2 p_1}e^{{\pi i\over 2}t_1 p_2}$
|$1$\cr
$\bf{Z_2\times Z_6}$|$e^{\pi it_2 p_1}e^{\pi it_1 p_2}$
|$e^{\pi it_2 p_1}$
\endtable

Several points are worth mentioning before we discuss the above
models as IIB orientifolds.  The irreducible anomalies vanish
for all of these models, and most can be interpreted as
Voisin-Borcea models \vois\borc .  It is interesting to note that a
continuation to negative $k=(r-a)/2$ of the Voisin-Borcea
classification elucidated by \eighto$\,$ applies to the $(3,51)$
if we choose $r=2$ and $a=4$. A similar remark can be made
for the other $J$ models.  The $\bf{Z_2\times Z_6}$
model with discrete torsion provides another F theory
realization of the Sen model \sen\gomu , and we will encounter this
same model as M theory on $\bf{T^5 /Z_6}$.  
Finally,
we are able to realize models in F theory sharing the moduli space of
the Gimon and Johnson models 
by applying the $J$ type torsion.  The $\bf{Z_2\times Z_2}$ case
is the regular discrete torsion yielding the $(3,51)$. One obtains
a couple of mirror orbifolds from this construction.  Notice from
Table 1 that the twisted sectors that will correspond to the
new type of twisted string are absent here, giving evidence that
it was correct for Gimon and Johnson to ignore these strings
in their models.  Whether there is a more systematic or
mathematical description of $J$ in the F theory context is not known
to me.  Also, whether tensionless strings play a role in these
models remains to be seen.

\subsec{IIB orientifolds}

{\it{Closed string results}}

The table below shows the discrete orientifold groups of the
IIB models corresponding to the F theory models.

\centerline{\bf{Table 3}:}
\smallskip
\centerline{IIB orientifolds}
\smallskip
\begintable
Model|Orientifold Generators\crthick
$\bf{Z_2\times Z_2}$|$(\Omega (-1)^{F_L}R_3,\Omega (-1)^{F_R}R_4)$\cr
$\bf{Z_4}$|$(\Omega (-1)^{F_L}R_3\alpha)$\cr
$\bf{Z_2\times Z_3}$|$(\Omega (-1)^{F_L}R_3,\omega ^2)$\cr
$\bf{Z_2\times Z_4}$|$(\Omega (-1)^{F_L}R_3,\alpha)$\cr
$\bf{Z_2\times Z_6}$|$(\Omega (-1)^{F_L}R_3, \omega)$
\endtable
\bigskip

Here, $(-1)^{F_L}=e^{2\pi is_1^L}$, $R_3 =e^{\pi is_3}$,
$R_4=e^{-\pi is_4}$, $\alpha =e^{{\pi i(s_3-s_4)\over 2}}$, and
$\omega =e^{{i\pi (s_3-s_4)\over 3}}$.  Again, the notations follow 
\bz$\,$ with $s_3=s_3^L +s_3 ^R$ and $s$ the spin.  Discrete torsion in
various twisted sectors is exactly as in the F theory 
description.  We will not discuss the $J$ models 
which should correspond to the Gimon and Johnson models.  The
first thing to notice about these models is that all cases
except the $\bf{Z_2\times Z_3}$ and $\bf{Z_4}$ cases contain
a $\bf{Z_2\times Z_2}$ subgroup.  Thus, the results of \bz$\,$
apply, and there will be $32$ each of the two kinds of
seven-branes with matter projected out in the $7-7'$ sector.
The matrices acting on Chan-Paton factors will be exactly
as in \bz$\,$ for the $\bf{Z_2\times Z_2}$ sectors.  Thus, what remains
is to determine the other twisted sectors.  Before doing this,
let us determine the closed string spectrum of the models.
The following tables show the general case and then specific
results from our models.  Let $g=e^{{2\pi i\over n}(s_3-s_4)}$
be the $\bf{Z_n}$ generator.
\vfil\eject
\bigskip
\centerline{\bf{Table 4}}
\centerline{Left-Moving States}

$$\matrix{{\rm Sector}&{\rm State}&(-1)^{F_L}R_3&g^m\cr\cr
{\rm untwisted}&&\cr\cr
{\rm NS}:&\psi^\mu_{-1/2}|0\!>&1&1\cr
&\psi^{z_3}_{-1/2}|0\!>&-1&e^{{2\pi im\over n}}\cr
&\psi^{\bar{z_3}}_{-1/2}|0\!>&-1&e^{{-2\pi im\over n}}\cr
&\psi^{z_4}_{-1/2}|0\!>&1&e^{{-2\pi im\over n}}\cr
&\psi^{\bar{z_4}}_{-1/2}|0\!>&1&e^{{2\pi im\over n}}\cr\cr
{\rm R}:&s_1^L=s_2^L,s_3^L=s_4^L={1\over 2}&-i&1\cr
&s_1^L=s_2^L,s_3^L=s_4^L={-1\over 2}&i&1\cr
&s_1^L=-s_2^L,s_3^L=-s_4^L={1\over 2}&-i&
e^{{2\pi im\over n}}\cr
&s_1^L=-s_2^L,s_3^L=-s_4^L={-1\over 2}&i&
e^{{-2\pi im\over n}}\cr\cr
{\rm twisted\,\, by}\,\, g^p\ne 1/2&&\cr\cr
{\rm NS}:&\psi^{z_3}_{p/n-1/2}|0\!>&e^{\pi i(1-{p\over n})}&
e^{2\pi i{m\over n}(1-{2p\over n})}\cr
&\psi^{\bar{z_4}}_{p/n-1/2}|0\!>&e^{-\pi i{p\over n}}&
e^{2\pi i{m\over n}(1-{2p\over n})}\cr\cr
{\rm R}:&s_1^L=-s_2^L&e^{\pi i({3\over 2}-{p\over n})}&
e^{2\pi i{m\over n}(1-{2p\over n})}\cr\cr
{\rm twisted\,\, by\, 1/2,\,\, n\,\, even}&&\cr\cr
{\rm NS}:&s_3^L=s_4^L={1\over 2}&i&1\cr
&s_3^L=s_4^L={-1\over 2}&-i&1\cr\cr
{\rm R}:&s_1^L=-s_2^L&-1&1\cr}$$
\vfill\eject
\bigskip
\centerline{Right-Moving States}

$$\matrix{{\rm Sector}&{\rm State}&(-1)^{F_L}R_3&g^m\cr\cr
{\rm untwisted}&&\cr\cr
{\rm NS}:&\psi^\mu_{-1/2}|0\!>&1&1\cr
&\psi^{z_3}_{-1/2}|0\!>&-1&e^{{2\pi im\over n}}\cr
&\psi^{\bar{z_3}}_{-1/2}|0\!>&-1&e^{{-2\pi im\over n}}\cr
&\psi^{z_4}_{-1/2}|0\!>&1&e^{{-2\pi im\over n}}\cr
&\psi^{\bar{z_4}}_{-1/2}|0\!>&1&e^{{2\pi im\over n}}\cr\cr
{\rm R}:&s_1^R=s_2^R,s_3^R=s_4^R={1\over 2}&i&1\cr
&s_1^R=s_2^R,s_3^R=s_4^R={-1\over 2}&-i&1\cr
&s_1^R=-s_2^R,s_3^R=-s_4^R={1\over 2}&i&
e^{{2\pi im\over n}}\cr
&s_1^R=-s_2^R,s_3^R=-s_4^R={-1\over 2}&-i&
e^{{-2\pi im\over n}}\cr\cr
{\rm twisted\,\, by}\,\, g^p\ne 1/2&&\cr\cr
{\rm NS}:&\psi^{z_3}_{p/n-1/2}|0\!>&e^{\pi i{p\over n}}&
e^{-2\pi i{m\over n}(1-{2p\over n})}\cr
&\psi^{\bar{z_4}}_{p/n-1/2}|0\!>&e^{-\pi i(1-{p\over n})}&
e^{-2\pi i{m\over n}(1-{2p\over n})}\cr\cr
{\rm R}:&s_1^R=-s_2^R&e^{-\pi i({1\over 2}-{p\over n})}&
e^{-2\pi i{m\over n}(1-{2p\over n})}\cr\cr
{\rm twisted\,\, by\, 1/2,\,\, n\,\, even}&&\cr\cr
{\rm NS}:&s_3^R=s_4^R={1\over 2}&i&1\cr
&s_3^R=s_4^R={-1\over 2}&-i&1\cr\cr
{\rm R}:&s_1^R=-s_2^R&1&1\cr}$$

\bigskip
Because we are using $\Omega$ not $\Omega J$, some of the phases are
different from that case.  As usual, $\Omega$ gives a positive
sign for symmetric left and right Neveu-Schwarz--Neveu-Schwarz
states but a negative sign for symmetric left and right 
Ramond-Ramond states.  In every model considered the untwisted
sector gives the $N=1$ supergravity multiplet.  In the twisted
sectors, the action of $g$ on the fixed points must be taken
into account, and similarly the discrete torsion projections
should be remembered.
\vfill\eject
\centerline{\bf{Table 5}:}
\smallskip
\centerline{Closed string sectors of IIB models}
\bigskip\bigskip
{\hsize=3in{
\noncenteredtables
\line{
\begintable\indent

Model|Twist|T|H\crthick
$\bf{Z_2\times Z_2}$|$1~$|$1~~$|$4~~$\cr
|$1/2$|$16$|$0$\crthick
$\bf{Z_4}$|$1$|$3$|$2$\cr
|$1/2$|$10$|$6$\crthick
$\bf{Z_2\times Z_3}$|$1$|$1$|$2$\cr
|$1/3$|$6$|$3$\cr
|$2/3$|$6$|$3$\crthick
$\bf{Z_2\times Z_4}$|$1$|$1$|$2$\cr
|$1/2$|$10$|$0$\cr
|$1/4$|$4$|$0$\cr
|$3/4$|$4$|$0$\crthick
$\bf{Z_2\times Z_6}$|$1$|$1$|$2$\cr
|$1/2$|$6$|$0$\cr
|$1/3$|$4$|$1$\cr
|$2/3$|$4$|$1$\cr
|$1/6$|$1$|$0$\cr
|$5/6$|$1$|$0$\cr
$\bf{Z_2\times Z_2}$,D.T.|$1~$|$1~$|$4~$\cr
|$1/2$|$0$|$16$\endtable
\hfil
\hbox{%
     \hbox to 0pt{\hskip-20pt\raise 150pt\hbox{
\begintable
$\bf{~~Z_2\times Z_4}$,D.T.~~|$~~1~$|$~~1~$|$~~2~$\cr
|$1/2$|$10$|$0$\cr
|$1/4$|$0$|$4$\cr
|$3/4$|$0$|$4$\crthick
$\bf{Z_2\times Z_6}$,D.T.|$~1~$|$~1~$|$~2~$\cr
|$1/2$|$0$|$6$\cr
|$1/3$|$4$|$1$\cr
|$2/3$|$4$|$1$\cr
|$1/6$|$0$|$1$\cr
|$5/6$|$0$|$1$
\endtable}}}
}
}
\bigskip
\hsize=6in{
\centerline{These results agree sector by sector with those}
\centerline{found in the last section from F theory.}
}
}
\vfill\eject

{\it{Tadpoles}}

The general formalism of 
Gimon and Polchinski \six$\,$ applies to the calculation of tadpole
anomalies in the above theories.  It is a bit too tedious to
show all of the theta functions that enter these calculations, but 
there are several points to keep in mind.  The projector
$\Omega (-1)^{F_L}R_3$ as compared to $\Omega$ gives an
extra minus sign on the closed vacuums twisted by $1\over 2$,
changing the sign of some Klein bottle and Mobius strip terms.
The untwisted Klein bottle gets no extra factor when one traces the 
$g$ action on the four-torus because the sum over fixed points
cancels the continuous sum(integral).  However, the $1\over 2$
twisted terms in the Klein bottle generally do get a fixed
point factor.  Also, the sum over fixed points of $R_3$ and
$R_4$ must be accompanied by appropriate phases for the action
of $g$ on the fixed points.  In fact, we will diagonalize the
action on combinations of fixed points.

Tadpoles twisted by $1\over 2$ for the $\bf{Z_2\times Z_2}$
case with discrete torsion get an overall minus with respect
to the $\bf{Z_2\times Z_2}$ without discrete torsion.  The
matrices operating on Chan-Paton factors are the same as for that
case, but the extra minus projects out the $SO(8)^8$ vectors
and keeps the global $SO(8)^8$ hypermultiplets.  The result agrees
with F theory.  Notice that if there were not a global $SO(8)^8$
symmetry, this model would be the same as the $(3,243)$
which is also the Gimon and Polchinski model at a generic 
point in the moduli space.  There appears to be no way to deform this
orientifold because moving branes away from the fixed points
would ruin the $SO(8)$ symmetry and cause the theory to be anomalous.

Tadpoles for the other models are listed as follows.  All are
proportional to $v_6 \int_0^{\infty} dl$ where 
$v_6=V_6/(4\pi\alpha ')^3$ and $V_6$ is the noncompact volume.

\eqn\zfour{{\bf{Z_4}}:{1\over 16}\sum_{I,I'}(\beta_I\beta_{I'} 16)^2}
\eqn\ztwozfour{{\bf{Z_2\times Z_4}}:\eqalign{&{1\over 32}
\sum_{I,I'}(\Tr
\gamma_{1/2}^I - \Tr\gamma_{1/2}^{I'} - \beta_I\beta_{I'}16)^2 \cr
&({1\over 32}
\sum_{I,I'}\alpha_I \alpha'_{I'}(\Tr
\gamma_{1/4}^I - \Tr\gamma_{1/4}^{I'})^2)+(1/4\rightarrow 3/4)\cr}}
\eqn\ztwozthree{{\bf{Z_2\times Z_3}}:({3\over 24}\sum_I \omega_I
(\Tr\gamma^I_{1/3} - \beta'_I 8)^2)+(1/3\rightarrow 2/3)}
\eqn\ztwozsix{{\bf{Z_2\times Z_6}}:\eqalign{&({3\over 48}\sum_{I,I'}
\omega'_I \omega^{''}_{I'} (\Tr\gamma^I_{1/3} - \Tr\gamma^{I'}_{1/3}
- \beta'_I 8 - \beta'_{I'} 8)^2)+(1/3\rightarrow 2/3)\cr
&({1\over 48}\sum_{I,I'}\omega^{' 2}_I \omega^{'' 2}_{I'}
 (\Tr\gamma_{1/6}^I
- \Tr\gamma_{1/6}^{I'})^2) + (1/6\rightarrow 5/6)\cr}}
\eqn\ztwozfourdt{{\bf{Z_2\times Z_4}}, D.T.:\eqalign{&{1\over 32}
\sum_{I,I'}(\Tr
\gamma_{1/2}^I - \Tr\gamma_{1/2}^{I'} - \beta_I\beta_{I'}16)^2 \cr
&(-{1\over 32}
\sum_{I,I'}\alpha_I \alpha'_{I'}(\Tr
\gamma_{1/4}^I - \Tr\gamma_{1/4}^{I'})^2)+(1/4\rightarrow 3/4)\cr}}
\eqn\ztwozsixdt{{\bf{Z_2\times Z_6}}, D.T.:\eqalign
{&({3\over 48}\sum_{I,I'}
\omega'_I \omega'_{I'} (\Tr\gamma^I_{1/3} - \Tr\gamma^{I'}_{1/3})^2)
+(1/3\rightarrow 2/3)\cr
&(-{1\over 48}\sum_{I,I'}\omega^{' 2}_I \omega^{' 2}_{I'} 
(\Tr\gamma_{1/6}^I
- \Tr\gamma_{1/6}^{I'})^2) + (1/6\rightarrow 5/6)\cr}}

\vskip 2in
The notations used in the above equations are defined below. 
All blocks are eight dimensional in the case of vectors and
$8\times 8$ dimensional in the case of matrices. In the above
$I$ runs over the fixed points of $R_3$ and $I'$ over those
of $R_4$.  In the solutions that follow, I have not been terribly
concerned about proving their uniqueness but have chosen the
phases to obtain the desired result that is consistent with
space-time anomaly cancellation.
\vfil\eject
\eqn\note{\eqalign{\alpha_I &=(1,-1,1,1)\cr
\alpha'_{I'}&=(-1,1,1,1)\cr
\omega_I&=(1,1,\omega^2 ,\omega^4)\cr
\omega'_I&=\omega_I\cr
\omega^{''}_{I'}&=(1,1,\omega^4 ,\omega^2) } }
\eqn\notem{ \eqalign{ \beta_I&=\pmatrix{0&&&\cr 
&0&&\cr 
&&1&\cr 
&&&1\cr }\cr
\beta'_I&=\pmatrix{ 1&&&\cr &0&&\cr &&0&\cr &&&0\cr }\cr
\gamma_{\Omega (-1)^{F_L}R_3} =\gamma_{1/2}^I&=
\pmatrix{ 1&&&\cr
&1&&\cr &&1&\cr &&&1\cr }\cr
\gamma^I_{1/4} &=\pmatrix{ 1&&&\cr &-1&&\cr &&1&\cr &&&1\cr }\cr
\gamma^I_{1/3} &=\pmatrix{ 1&&&\cr &1&&\cr &&\omega^2 &\cr &&&
\omega^4\cr }\cr
\gamma^I_{1/6} &=\pmatrix{ 1&&&\cr &1&&\cr &&\omega^4 &\cr 
&&&\omega^2\cr }}}

\vskip 1in
These matrices imply that D-branes yield the following matter 
content and left-over tadpole anomaly.  The anomaly is 
multiplied by $v_6\int_0^{\infty} dl$ in the following table
and the expected extra vectors from F theory are listed in the 
last column.
\vfil\eject

\centerline{{\bf{Table 6}}:}
\smallskip
\centerline{Tadpole Anomalies and Extra Gauge Fields from F Theory}
\smallskip
\begintable
Model|V|H|Anomaly|Extra Vectors\crthick
${\bf{Z_2\times Z_2}}$|$SO(8)^8$|$0$|$0$|$0$\cr
${\bf{Z_4}}$|$0$|$0$|$64$|$SO(8)^4$\cr
${\bf{Z_2\times Z_3}}$|$SO(8)^2$|$SO(8)$|$48$|$SO(8)^3$\cr
${\bf{Z_2\times Z_4}}$|$SO(8)^6$|$0$|$64$|$SO(8)^4$\cr
${\bf{Z_2\times Z_6}}$|$SO(8)^4$|$0$|$64$|$SO(8)^4$\cr
${\bf{Z_2\times Z_2}}$, D.T.|$0$|$SO(8)^8$|$0$|$0$\cr
${\bf{Z_2\times Z_4}}$, D.T.|$0$|$SO(8)^2$|$0$|$0$\cr
${\bf{Z_2\times Z_6}}$, D.T.|$0$|$SO(8)^2$|$32$|$SO(8)^2$
\endtable
\bigskip

The results are again in agreement with F theory as the left-over 
tadpole anomaly is proportional to the irreducible anomaly, to
the number of $SO(8)$ vectors necessary to cancel this anomaly,
and to the expected extra vectors from F theory.  We, thus,
assume that there are twisted open strings present in these
theories and will provide other evidence for these objects in the
M theory discussion.

\newsec{Orientifolds from M Theory}

\subsec{T-duality, S-duality, and IIB orientifolds}

In this section two M theory orbifolds, $\bf{T^5/Z_4}$ and
$\bf{T^5/Z_6}$, will be studied.  The $\bf{Z_4}$ generator
is $(-,\alpha)$ and the $\bf{Z_6}$ generator $(-,\omega)$ where
the minus acts on the fifth circle while $\alpha$ or $\omega$
is acting on the four-torus.  We will first give an argument
for why these models should be equivalent to the $(K3\times 
{\bf{S^1}})/{\bf{Z_2}}$ first discussed by \sen .  Writing $\bf{Z_4}$
($\bf{Z_6}$) as an internal direct product $\bf{Z_4/Z_2\times Z_2}$
($\bf{Z_6/Z_3\times Z_3}$), we see that the two models are
equivalent to $(K3\times {\bf{S^1}})/{\bf{Z_2}}$.  This extra 
$\bf{Z_2}$
acts in both cases with a $-1$ on $\bf{S^1}$ and eight $-1$'s
on the $19$ antiself-dual two-forms of $K3$.  The action is not
the same as in \sen , for only some of the $-1$'s come from
exchanging part of the two $E_8$ lattices.  Presumably, there is
an $SO(19)$ rotation that relates the three $\bf{Z_2}$'s.  In
any case the spectrum of the three models using the argument
of \sen$\,$ is $9$ tensors, $20$ hypermultiplets, and a gauge
group of rank eight with hypermultiplets in the adjoint of
this gauge group.  This spectrum also corresponds to F theory
on the $(19,19)$.

Using the prescription of \sentwo$\,$  to translate the models into
IIA theory, we get the orbifolds $(-1)^{F_L}\alpha$ and
$(-1)^{F_L}\omega$.  The massless spectrum of IIA and IIB
on these orbifolds is given in the following table.

\centerline{{\bf{Table 7}}:}
\smallskip
\centerline{Type II Strings on $(-1)^{F_L}\alpha$ and 
$(-1)^{F_L}\omega$}
\smallskip
\begintable
Model|Sector|T|V|H\crthick
IIA on ${\bf{T^4}}/(-1)^{F_L}\alpha$|1|1|2|2\cr
|$\alpha^2$|0|6|10\cr
|$(-1)^{F_L}\alpha$|4|0|4\cr
|$(-1)^{F_L}\alpha^3$|4|0|4\crthick
IIA on ${\bf{T^4}}/(-1)^{F_L}\omega$|1|1|0|2\cr
|$\omega^2$|0|4|5\cr
|$\omega^4$|0|4|5\cr
|$(-1)^{F_L}\omega^3$|6|0|6\cr
|$(-1)^{F_L}\omega$|1|0|1\cr
|$(-1)^{F_L}\omega^5$|1|0|1\crthick
IIB on ${\bf{T^4}}/(-1)^{F_L}\alpha$|1|3|0|2\cr
|$\alpha^2$|6|0|10\cr
|$(-1)^{F_L}\alpha$|0|4|4\cr
|$(-1)^{F_L}\alpha^3$|0|4|4\crthick
IIB on ${\bf{T^4}}/(-1)^{F_L}\omega$|1|1|0|2\cr
|$\omega^2$|4|0|5\cr
|$\omega^4$|4|0|5\cr
|$(-1)^{F_L}\omega^3$|0|6|6\cr
|$(-1)^{F_L}\omega$|0|1|1\cr
|$(-1)^{F_L}\omega^5$|0|1|1
\endtable
\vfil\eject

Notice that if one excludes one tensor from the untwisted sectors, the 
IIB models are obtained from the IIA models by exchanging vectors
and tensors.  Because the two models are the same, this exchange
is a symmetry (as long as the gauge group is abelian).  This
result is expected since the spectrum matches the $(19,19)$
which is mirror symmetric.  If we compactify M theory on a
circle, we obtain the orientifolds of the IIA theory on
$\bf{T^5/Z_4}$ and $\bf{T^5/Z_6}$.  Applying T-duality on the
fifth circle gives IIB on the orientifolds 
$({\bf{T^4}}/{\Omega\alpha})
\times{\bf{S^1}}$ and  $({\bf{T^4}}/{\Omega\omega})\times{\bf{S^1}}$.
The expected spectrum of these models is clearly
different from the results of Gimon and Johnson reduced to five
dimensions.\foot{These T-dual models were obtained in discussions with
S. Mukhi.}  The expected massless spectrum agrees with the 
$(-1)^{F_L}$ models as S-duality arguments would predict.
However, the expected results require again twisted open strings 
which we discuss in the next section.

\subsec{Twisted Strings}

The first piece of evidence for twisted strings will come from
comparing the Klein bottle term of the $\Omega{\bf{Z_4}}$ model
to the corresponding sectors of the partition function of the 
$(-1)^{F_L}{\bf{Z_4}}$ model.  Using the notations of \six$\,$
the Klein bottle amplitude in the loop formulation is

\eqn\omegaalpha{(1-1){v_6\over 16}\int_0^{\infty}{dt\over t^4}
(8 {f_3^4(e^{-2\pi t})f_4^4(e^{-2\pi t})\over 
f_1^4(e^{-2\pi t})f_2^4(e^{-2\pi t})} - 
8 {f_2^4(e^{-2\pi t})f_4^4(e^{-2\pi t})\over 
f_1^4(e^{-2\pi t})f_3^4(e^{-2\pi t})}).}

By doing a duality transformation $t\rightarrow {1\over t}$, one
sees that the divergence vanishes so this would appear to
be a consistent closed string theory as noted by \eight .

Corresponding to the Klein bottle, we examine the following
sectors $Z(t_{\sigma},t_{\tau})$ of the $(-1)^{F_L}\alpha$ toroidal
partition function:$Z(1,(-1)^{F_L}\alpha)+ 
Z(1,(-1)^{F_L}\alpha^3)+Z(\alpha^2 ,(-1)^{F_L}\alpha)+ 
Z(\alpha^2 ,(-1)^{F_L}\alpha^3)$.
The modulus of the torus is set to be $\tau =it$ and $t_{\sigma}$
($t_{\tau}$) is the twist in the spatial(time) direction of the 
torus.  This gives

\eqn\theta{\eqalign{ &4{\left(\Theta^2{0\choose 0}
\Theta{0\choose 1/4}
\Theta{0\choose 0 } - \Theta^2{0\choose 1/2 }
\Theta{0\choose
3/4 }\Theta{0\choose -3/4 }\right)^2 - \Theta^4
{1/2\choose 0 }
\Theta^2{1/2\choose
1/4 }\Theta^2{1/2\choose -1/4 } \over
\eta^{12}\Theta^2{1/2\choose
1/4 }\Theta^2{1/2\choose -1/4 } }\cr
&- {\left(\Theta^2{0\choose 0 }\Theta{1/2\choose
1/4 }\Theta{1/2\choose -1/4} - \Theta^2{0\choose 
1/2 }
\Theta{1/2\choose
3/4 }\Theta{1/2\choose -3/4 }\right)^2 + \Theta^4
{1/2\choose 0 }
\Theta^2{0\choose
1/4 }}\Theta^2{0\choose -1/4 } \over
\eta^{12}\Theta^2{0\choose
1/4 }\Theta^2{0\choose -1/4 } }.} 

Here the notations are from the appendix of \orb$\,$ with 
$q=e^{-2\pi t}$.
Using
identities found in \eight , the above reduces to

\eqn\thetatwo{(1-1)(8 {f_3^4(e^{-2\pi t})f_4^4(e^{-2\pi t})\over 
\eta^8(e^{-2\pi t})f_2^4(e^{-2\pi t})} - 
8 {f_2^4(e^{-2\pi t})f_4^4(e^{-2\pi t})\over 
\eta^8(e^{-2\pi t})f_3^4(e^{-2\pi t})})=
(1-1)8{f_4^2\over \eta^8 f_2^4 f_3^4}.}

This matches the Klein bottle term if one replaces $\eta^8$ by
$f_1^4$.  Since S-duality mixes up the R-R and NS-NS sectors,
the $(1-1)$ here does not correspond to NS-NS$-$R-R.  The
factor $f_1^4$ or $\eta^8$ represents the bosonic space-time 
oscillators and only affects the massive spectrum.  Perhaps,
this difference between the two theories reflects a
renormalization and would be absent if perturbative and
nonperturbative corrections were taken into account.  The
difference can be eliminated if it is sensible in a perturbative
framework to define an operator $\Omega_b$ which exchanges 
left and right-moving, space-time bosonic oscillators.  Then S-duality
would exchange $\Omega$ and $(-1)^{F_L}\Omega_b$.  In order
for the $(-1)^{F_L}\alpha$ theory to be consistent and 
modular invariant, strings twisted by $(-1)^{F_L}\alpha$
are required.  At the modulus $\tau_1=0$, the above sectors of
the partition function are invariant under $\tau\rightarrow
\tau +1$ so the modular transformation that produces these
strings corresponds to $t\rightarrow{1\over t}$ as with the Klein 
bottle.  Since there is no phase transition at $\tau_1=0$, even
at $t=0$ (The only dangerous term is the massless contribution
which vanishes.), one would not expect these strings to 
disappear there.  Thus, the $t\rightarrow{1\over t}$ transformation
should also produce these strings in the $\Omega$ theory.
Since there is no space-time anomaly, their contribution to 
the Klein bottle vanishes.

The second piece of evidence for these strings is to provide a
perturbative formulation of them.  If we try to solve for a
string twisted by $\Omega\alpha$, we obtain an open string
stuck at the fixed point with no zero or oscillator modes
on the four-torus.  Thus, the vacuum energy of the Ramond sector 
is $0$ and that of the Neveu-Schwarz sector is $-1/4$ so these
sectors do not match and supersymmetry is violated.  These
strings must somehow be projected out of the spectrum.  The
hypothesis is that $J$ is the required projection and that
twisting with $\Omega$ corresponds to projecting with $\Omega J $
and vice versa.  In the F theory version of $J$, one might
postulate that a transition to the smooth Calabi-Yau is impossible 
and that blowing up the fixed point might create a 
superpotential along the lines of \supone\suptwo .
On the other hand, if we try to solve the
equation $Z_3(\sigma +\pi)=\Omega J\alpha Z_3(\sigma)=
\Omega\alpha Z_3^c(\sigma)$, we do get a viable open string.
Here, $J$ is the operator that switches the twist from
$1/4$ to $3/4$ and if $Z_3$ is twisted by $1/4$, $Z_3^c$ is
twisted by $3/4$.  The solution for $Z_3$ is the following:

\eqn\pert{Z_3=z^{fix}_3 +\sum_n(\alpha^3_{-n-1/4}e^{-i(n+{1\over 4})
2(\tau - \sigma)}+\tilde\alpha^3_{-n-3/4}e^{-i(n+{3\over 4})
2(\tau + \sigma)})}

and

\eqn\pertc{Z_3^c=z^{fix}_3 +\sum_n (i\alpha^3_{-n-1/4}
e^{-i(n+{1\over 4})
2(\tau + \sigma)}+i\tilde\alpha^3_{-n-3/4}e^{-i(n+{3\over 4})
2(\tau - \sigma)})}

Since the orbifold has $z_3\sim iz_3$, the boundary conditions
can be such that $({\partial z_3\over\partial\sigma}
{\partial z_3\over\partial\tau})_{\sigma=0}=
-({\partial z_3\over\partial\sigma}
{\partial z_3\over\partial\tau})_{\sigma=\pi}$ as they are here.
Notice also that $Z_3^c=\Omega Z_3$.  Trying to twist the
NS and R operators acting on the vacuum by $\Omega J\alpha$
appears to lead to inconsistencies because of the extra phases
of the twist operators unless one assumes that the action of
$\Omega$ in these twisted sectors also reverses the GSO
projection.  This affect would be similar to twisting by
$(-1)^{F_L}$.  In order to have an invariant combination under
the $\Omega\alpha$ projection, these strings must always 
occur in pairs, one string and its orientation reversed counterpart.
If one assumes that $L_0^{string 1}=L_0^{string 2}$ where $L_0$
is the Virasoro generator, the massless spectrum is identical
to the $(-1)^{F_L}\alpha$ twisted string.  These two strings
would result from pulling apart the left and right moving
sectors of the $(-1)^{F_L}\alpha$ twisted string to create
two strings.  At the massive level, to produce complete
agreement between the two theories probably requires modifying
$(-1)^{F_L}$ to $(-1)^{F_L}\Omega_b$ or taking into account
higher order corrections to the spectrum as discussed above.
The spectrum of these twisted strings generally contains a
massless vector and a massless hypermultiplet, but 
how the hypermultiplets
are projected out and the gauge group becomes nonabelian in
some F theory models will not be resolved here.

\newsec{Discussion}

This paper has probably raised more questions than it has resolved.
We have considered orbifolds of F theory having a $\bf{Z_2}$
singularity on the fiber which allows for a perturbative IIB
description.  It is natural to wonder whether F theory orbifolds
with a $\bf{Z_n}$ singularity, $n>2$, that are stuck at strong
coupling can be understood in any perturbative framework.
The affect of discrete torsion in F theory has been seen
to turn gauge symmetry into global symmetry.  The significance
of this result could probably be better understood.  We have
considered the $J$ torsion in F theory but not understood what
$J$ corresponds to geometrically and whether tensionless strings
play a role.  The orientifolds we have discussed have required
a new type of twisted string that has a perturbative description.
How this string fits into the D-brane framework and how a
number of these strings come together to yield a nonabelian
gauge symmetry is an open question. We have observed that the
objects in the $\Omega{\bf{Z_4}}$ theory are ``meson''-like, 
but one could conjecture that there are possible ``baryon''-like
objects formed from twisted strings.  We have also found that
$(-1)^{F_L}$ should be modified, at least perturbatively, to
have S-duality valid for the entire spectrum.  Many of the models
analyzed here have been realized in seemingly different 
contexts such as the M theory orbifolds that are not readily
converted into their F theory counterparts because of the
nontrivial action on the twelth dimension.
One is therefore led to speculate about some underlying unified 
description.

\bigskip\centerline{\bf Acknowledgments}\nobreak

I thank K. Dienes, C. Johnson, S. Mukhi, A. Sen, 
and E. Witten for useful discussions
and especially A. Zaffaroni for fruitful collaboration and
P. Bozzay for help preparing the tables.  This research was 
supported in part by NSF Grant PHY-9513835.

\listrefs
\end